# Transverse Beam Shape Measurements of Intense Proton Beams Using Optical Transition Radiation

Victor E. Scarpine[a]

[a]*Fermilab, PO Box 500, Batavia, IL 60510 USA*

**Abstract**

A number of particle physics experiments are being proposed as part of the Department of Energy HEP Intensity Frontier. Many of these experiments will utilize megawatt level proton beams onto targets to form secondary beams of muons, kaons and neutrinos. These experiments require transverse size measurements of the incident proton beam onto target for each beam spill. Because of the high power levels, most beam intercepting profiling techniques will not work at full beam intensity. The possibility of utilizing optical transition radiation (OTR) for high intensity proton beam profiling is discussed. In addition, previous measurements of OTR beam profiles from the NuMI beamline are presented.

*Keywords:* Type your keywords here, separated by semicolons ;  optical transition radiation

## 1. Introduction

Particle-beam diagnostic techniques based on optical transition radiation (OTR) have been demonstrated at a number of facilities over a wide range of beam energy (or Lorentz factor, γ) [1] - [3]. Optical transition radiation is generated when a charged particle transits the interface between two media with different dielectric constants [2]. This radiation is emitted over the visible spectrum; so, optical imaging techniques can be used to acquire the OTR signal and then reconstruct beam size and position at the dielectric surface. Since OTR is a surface phenomenon, thin foils are used as the converter to reduce beam scattering and minimize heat deposition.

Fermilab has developed OTR detectors for the Tevatron collider (Run II) and for high-intensity proton beams for neutrino experiments (NuMI). In addition, other laboratories such as CERN [4] and J-PARC [5] have built OTR detectors for measurements of intense proton beams. OTR detectors can be used to measure such beam properties as transverse profiles and 2-D shape, transverse position, divergence, emittance and intensity. The Fermilab OTR detectors are installed in a number of locations in the beam complex including:

- The A150 transport line between the Main Injector and the Tevatron to measure 150 GeV protons and antiprotons.
- The Tevatron ring for injection study measurements of 150 GeV protons out of the A150 transport line. This detector is adjacent to the Tevatron Ionization Profile Monitor (IPM).
- The NuMI transport line just upstream of the primary target to measure high-intensity 120 GeV protons.

These three locations require the OTR detector to operate over a large range of beam conditions. Table 1 gives a summary of the relevant beam conditions for our OTR detector design. This paper will focus on the operation of the OTR detector installed in the high-intensity NuMI beamline.

Table 1. Beam Parameters for OTR Detector Locations

|  | A150 Beamline | Tevatron | NuMI Beamline |
| --- | --- | --- | --- |
| Beam Type | Proton/Antiproton | Proton/Antiproton | Proton |
| Beam Energy (GeV) | 150 | 150 | 120 |
| Beam Pulse Intensity (E10) | ~ 1 to 50 | ~ 1 to 50 | ~ 10 to 5000 |
| Transverse Beam Size ($\sigma$ in mm) | ~ 1 to 4 | `1.5 | ~1.0 |
| Pulse Repetition Rate | ~ 10 per day | Studies only | ~0.5 Hz |

## 2. Fermilab OTR Detector and NuMI Beamline Installation

*2.1. OTR Detector Design*

A standard OTR detector design, based on initial beam measurements of high-intensity proton beams at Fermilab [6], has developed for use in various proton beamlines [7]. The Fermilab OTR design is based on collecting reverse OTR in order to make beam measurements. This is a standard technique that allows for the maximum collection of OTR. A discussion of a detector based on collecting forward OTR is presented later in this paper.

In general, the Fermilab OTR design utilizes opposing OTR foils to allow for bidirectional beam measurements. However, the NuMI OTR detector uses two aluminized Kapton foils to make unidirectional beam measurements. The foils are tilted 45 degrees to the beam axis in order to reflect the backward OTR perpendicular to the beamline. One foil is used as a primary OTR screen while the second foil is periodically inserted into the beam to track changes to the primary foil. The OTR detector also utilizes a radiation-hardened CID camera [8] and neutral density filters to allow operation over a range of beam intensities and under various radiation environments.

The NuMI OTR detector acquires beam profile images for every beam pulse. A front-end PC uses a Labview acquisition and analysis program to process these images. The analysis produces the beam X and Y centroids and sigmas as well as the beam tilt and ellipticity. A more detailed description of the OTR detector can be found in reference [7].

*2.2. NuMI Beamline*

The NuMI beamline [9] was initially designed to operate at $2 \times 10^{13}$ to $4 \times 10^{13}$ 120 GeV protons per pulse at a repetition rate of ~0.5 hertz, producing a beam power of up to 320 kW. Future upgrades to the beamline will increase the protons per pulse to ~$5 \times 10^{13}$ with a beam power of up to ~700 kW. The rms transverse beam size is ~ 1 mm at the NuMI target. Presently, NuMI uses a segmented titanium foil SEM just upstream of the NuMI target and shield wall as a target beam profile monitor [10]. This target SEM measures the beam profile for every pulse. The NuMI OTR detector is installed just downstream of the target SEM but still in front of the shield wall, thus allowing for comparison of beam measurements between the two devices.

## 3. NuMI OTR Results

Figure 1a shows OTR images for bunch intensities of ~$2.4 \times 10^{13}$ and ~$4.1 \times 10^{13}$. This figure also shows the beam projections with Gaussian fits. The images show the increase in beam size with an increase in beam intensity. The images also show an increase in the beam ellipticity with higher intensities. This shows the advantage of a two-dimensional beam shape monitor, such as an OTR detector, over other standard one-dimensional profile monitors.

The choice of an aluminized Kapton foil for the NuMI OTR was made to minimize the scatter of the NuMI beam. Issues of foil lifetime at this high beam intensity are not well understood, but it was assumed that the foil would change over time. Figure 1b shows a comparison of the measured horizontal beam σ from the SEM monitor and the OTR detector over a period of 80 days. One can see from the figure that the value of σ for the OTR detector is slowly diverging from the value from the SEM monitor. Although there is a difference in these two measurements, it is not clear that this could be attributed to a change in the OTR foil.

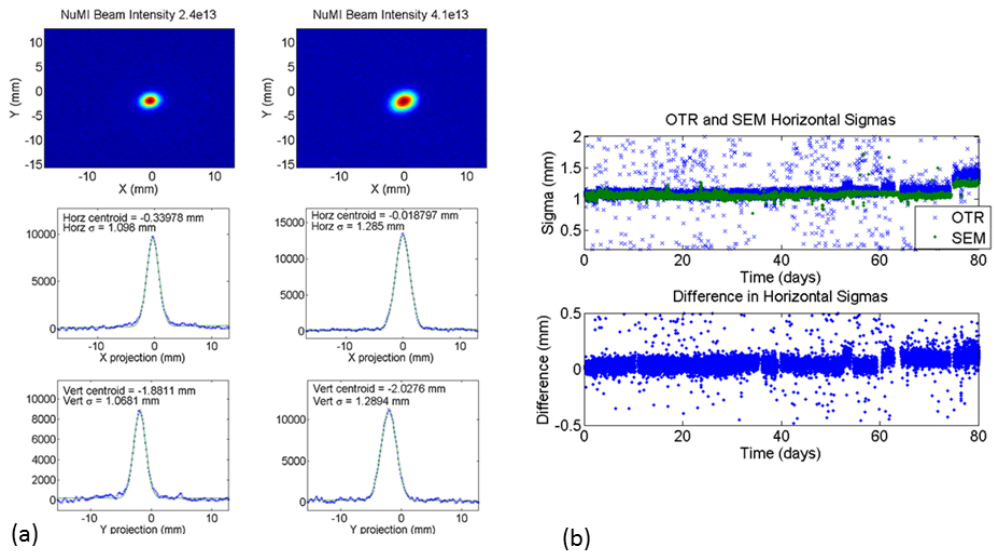

Fig. 1. (a) NuMI OTR beam images and horizontal and vertical projections with Gaussian fits for $2.4 \times 10^{13}$ and $4.1 \times 10^{13}$ protons per bunch; (b) comparison of the horizontal σ between the SEM monitor and the OTR detector over time.

## 4. OTR Foil Aging

One issue with OTR detectors for high-intensity proton beams is the change in the characteristics of the foil. Reverse OTR imaging is sensitive to the reflectivity and flatness of the intercepting foil. Figure 1b seems to indicate that foil aging has occurred with the NuMI detector. To check for aging of the primary foil, the secondary foil was temporarily inserted into the beamline. The OTR images from this foil were compared to OTR images taken with the primary foil to look for signs of change in the primary foil. Figure 2 shows OTR images and projections for the primary and secondary foils under similar beam conditions. These images give similar centroid and σ values for both foils, the OTR intensity from the primary foil is clearly reduced.

After exposure of the primary aluminized Kapton foil to ~$6.5 \times 10^{13}$ 120 GeV protons, the primary foil was removed from the detector. Figure 3 shows the front and back of the primary foil and illustrates both mechanical deformation and clouding of the aluminium reflective surface.

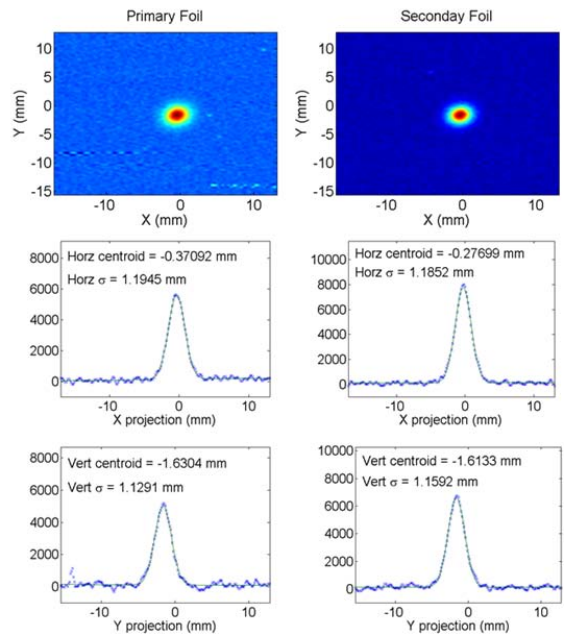

Fig. 2. Comparison of OTR images taken with the primary and the secondary foils under similar beam conditions after ~70 days of primary foil operation.

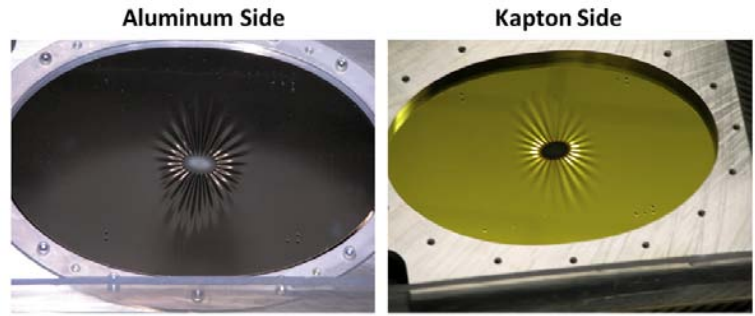

Fig. 3. Front (aluminum) and back (Kapton) side of primary foil after exposure to ~$6.5 \times 10^{13}$ 120 GeV protons.

## 5. Possible Forward OTR Imaging of Intense Beams

As shown above, exposure of OTR foils to prolonged high-intensity beams of protons can cause changes in the material properties of the foils, such as reflectivity and flatness. These material changes directly affect the ability of the OTR detector to generate reliable beam measurements. Since forward OTR is independent of the foil reflectivity and flatness, one possible method to mitigate these effects would be to make beam profile measurements utilizing forward OTR. Figure 4a shows a false color forward OTR angular emission pattern measurement for 120 GeV protons while figure 4b shows a slice profile through the center of the emission pattern.

Figure 4c shows a possible forward OTR imaging system to make beam measurements. This design uses a pickoff mirror downstream of a normal incident foil to collect a portion of the forward OTR light. In addition, the beam intercepting foil could be a beam vacuum window, which is already present in many beamlines. A series of focusing elements are then used to generate a beam profile image on to a camera. The light collected in this manor is much less than from standard reverse OTR detectors. However, for high-intensity beams, such as NuMI, the amount of OTR produced is adequate for image detection and measurement.

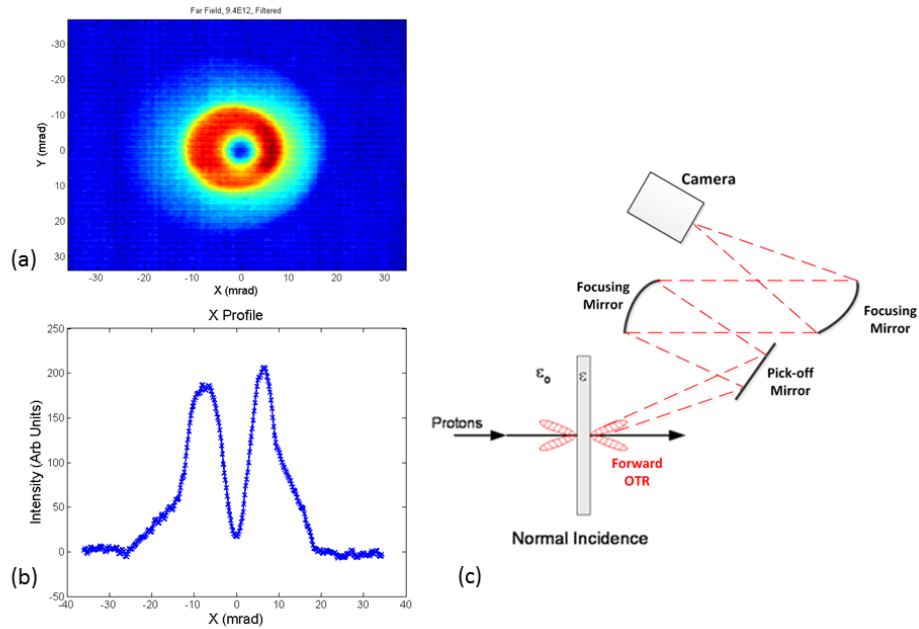

Fig. 4. (a) Forward OTR emission pattern; (b) forward OTR profile through emission center point; (c) a possible imaging system to collect forward OTR;

## 6. Conclusion

The use of OTR for a beam profiling system of high-intensity protons has been demonstrated at the Fermilab NuMI beamline. The OTR detector was able to produce beam position and size measurements for almost every beam pulse. However, prolonged exposure of the foil to beam, induced material changes to the 6 μm aluminized Kapton foil, which resulted in inaccurate measurements. A beam profiling detector based on forward OTR would eliminate many of these issues.

**Acknowledgements**

We thank our NuMI colleagues for their continued help and interest in this detector: Sam Childress, Peter Lucas, Gordon Koizumi, Jim Hylen and Doug Jenson. We also would like to thank the FNAL Accelerator Division for their contributions: Alex Lumpkin, Gianni Tassotto, Dave Slimmer, Dan Schoo, Rick Pierce, Linda Purcell-Taylor, Terry Anderson, Brad Tennis and Jim Fitzgerald. In addition, we wish to thank Carl Lindenmeyer, Jerry Zimmerman, John

Korienek, Ron Miksa, Karen Kephart, Eileen Hahn and Wanda Newby of FNAL Particle Physics Division for their outstanding work in the design and construction of the detector. We would also like to thank Bob Webber and Manfred Wendt for their continued encouragement.